\documentclass[conference]{IEEEtran}

\usepackage{amsmath, amssymb, amsfonts}
\usepackage{xcolor}
\usepackage{makecell}   
\usepackage{multirow}
\usepackage{pifont}
\usepackage{comment}
\usepackage{algorithm}
\usepackage[noend]{algorithmic}
\usepackage{subcaption}
\usepackage{balance}
\usepackage{graphicx}
\usepackage{url}
\usepackage[svgnames]{xcolor}
\usepackage{soul}
\usepackage{hyperref}

\sethlcolor{LightBlue}
\begin{document}

\title{FlexViT: A Flexible FPGA-based Accelerator for \\ Edge Vision Transformers}


\author{Hubert Dymarkowski, Xingjian Fu, Rappy Saha, Jude Haris, Jos\'e Cano \\
\emph{School of Computing Science, University of Glasgow, Scotland, UK}
}

\maketitle


\begin{abstract}

Deploying Vision Transformer (ViT) models on edge platforms remains challenging due to their high computational demands and the architectural heterogeneity of modern hybrid ViT models, which incorporate both fully connected and convolutional layers. 
This heterogeneity leads to significant variation in tensor shapes, requiring flexible and efficient FPGA-based acceleration. 
In this paper, we present FlexViT, a reconfigurable FPGA accelerator for efficient ViT inference on resource-constrained edge devices. Built on the SECDA-TFLite framework, FlexViT employs a hardware–software co-design approach that maps both fully connected and convolutional layers onto a unified high-throughput INT8 GEMM engine using a runtime \textit{im2col} transformation. 
To efficiently support diverse layer configurations, we propose a dual-mode dataflow that dynamically switches between input and weight reuse by reconfiguring the compute array at runtime. 
We further introduce a depth-first tiling strategy that completes accumulation in a single pass, eliminating off-chip partial-sum transfers and reducing memory bandwidth requirements. We implement FlexViT on a PYNQ-Z2 FPGA and evaluate it across a representative set of ViT models. 
FlexViT achieves up to 2.74$\times$ speedup on accelerator-executed layers, translating into up to 1.40$\times$ end-to-end speedup compared to CPU-only execution. 
The code is available at: \url{https://github.com/gicLAB/FlexViT}

\end{abstract}



\section{Introduction}
\label{sec:intro}

The rapid evolution of intelligent edge devices has increased the demand for on-device processing of complex visual data, avoiding the latency and bandwidth limitations of cloud-centric solutions. Field-Programmable Gate Arrays (FPGAs) are a compelling platform for such workloads due to their fine-grained parallelism and reconfigurability, enabling customized datapaths that improve latency and energy efficiency for computer vision (CV) applications. 

However, deploying modern Vision Transformers (ViTs) on edge platforms remains challenging~\cite{gibsonDLAS2025}. ViTs achieve superior performance in tasks such as detection and segmentation by capturing long-range dependencies, but their high computational complexity and memory requirements make them difficult to deploy on resource-constrained devices.

\begin{table}[t]
\small
\centering
\caption{Comparison of ViT Accelerators.}
\label{tab:sota_comparison}
\newcommand{\cmark}{\ding{51}}
\newcommand{\xmark}{\ding{55}}
\begin{tabular}{|l|ccc|}
\hline
\textbf{Paper} & \textbf{\shortstack{Standard \\ ViT}} & \textbf{\shortstack{Hybrid \\ ViT}} & \textbf{\shortstack{Open \\ Source}}  \\ \hline
Nag et al.~\cite{vita}  & \cmark & \xmark & \xmark \\
Shao et al.~\cite{10557992} & \xmark & \cmark & \xmark  \\
Cao et al.~\cite{an_energy_efficient} & \cmark & \xmark & \xmark \\
Liang et al.~\cite{m2vit}& \xmark & \cmark  & \xmark \\
\textbf{Ours (FlexViT)} & \cmark & \cmark & \cmark \\\hline
\end{tabular}
\end{table}

Furthermore, modern edge-oriented ViT models exhibit significant architectural diversity. Early compact models, such as Vision Transformer Tiny (ViT-T) and Data-efficient Image Transformers Tiny (DeiT-T), are dominated by fully connected (FC) operations, whereas recent hybrid architectures, such as MobileViT and EfficientViT, incorporate both convolutional (CONV) and FC components. 
This diversity creates a challenge for FPGA accelerator design: architectures optimized for a single computational pattern often suffer from poor generality as ViT models continue to evolve. Existing FPGA accelerators typically target either FC-dominated or convolution-dominated workloads, leaving a gap in flexible designs capable of supporting the full spectrum of modern ViT architectures, as summarized in Table~\ref{tab:sota_comparison}.

To address this diverse and evolving landscape, we identify three key challenges for efficient edge deployment:

\begin{itemize}
    \item \textbf{Layer Heterogeneity:} hybrid ViT models combine FC and CONV layers, each with distinct dataflows and memory access patterns, complicating efficient hardware support.

    \item \textbf{Dimensional Variance:} layer dimensions vary significantly across and within models, causing underutilization in monolithic accelerators when processing irregular matrix shapes (e.g., short–wide or tall–narrow tensors).
    
    \item \textbf{Resource Constraints:} edge FPGA systems-on-chip (SoCs) provide limited block RAM (BRAM) and digital signal processor (DSP) resources compared with data-center-class devices. Existing ViT accelerators often rely on large on-chip buffers, making them unsuitable for resource-constrained platforms.
\end{itemize}

In this paper, we address these challenges by proposing \textbf{FlexViT}, a flexible FPGA-based accelerator for edge devices that efficiently supports both standard and hybrid ViT models using the SECDA-TFLite toolkit~\cite{Haris2023JPDC}. FlexViT adopts a hardware–software co-design approach to address the following key features:

\begin{itemize}
    \item \textbf{Layer heterogeneity}: it employs a unified computation strategy that maps FC and CONV layers onto a single General Matrix–Matrix Multiplication (GEMM) engine. 

   \item \textbf{Dimensional variance}: it dynamically reconfigures the compute array to exploit input or weight reuse, improving utilization for irregular workloads.

    \item \textbf{Resource constraints}: it combines streaming execution with depth-first tiling to enable single-pass accumulation and reduce memory bandwidth requirements.
\end{itemize}

The main contributions of this work are as follows:

\begin{itemize}
    \item A systematic analysis of modern edge-oriented ViT models, highlighting key challenges for efficient acceleration on resource-constrained FPGA platforms.

    \item \textbf{FlexViT}, a flexible FPGA-based accelerator for Vision Transformers capable of efficiently supporting both standard and hybrid ViT architectures through a unified and reconfigurable design.
    
    \item A comprehensive evaluation on a PYNQ-Z2 platform, demonstrating $1.77\times$ to $2.74\times$ speedup on accelerator-offloaded 8-bit integer (INT8) quantized FC and CONV layers, and up to $1.40\times$ end-to-end latency improvement compared to CPU-only execution across five representative ViT models.
\end{itemize}

\section{Background}
\label{sec:background}

\subsection{Standard Vision Transformers (ViTs)}

Dosovitskiy~\textit{et al.}~\cite{dosovitskiy_image_2021} introduced the idea of using the transformer architecture~\cite{vaswani_attention_2023} on sequences of image patches~\cite{dosovitskiy_image_2021}, so they can be used in CV tasks such as image recognition and object detection. ViT models not only outperform many state-of-the-art CNNs, their direct competitors, but also require substantially less data and computational time to train~\cite{DeiT}.

\begin{figure}[t]
    \centering
    \includegraphics[width=0.85\linewidth]{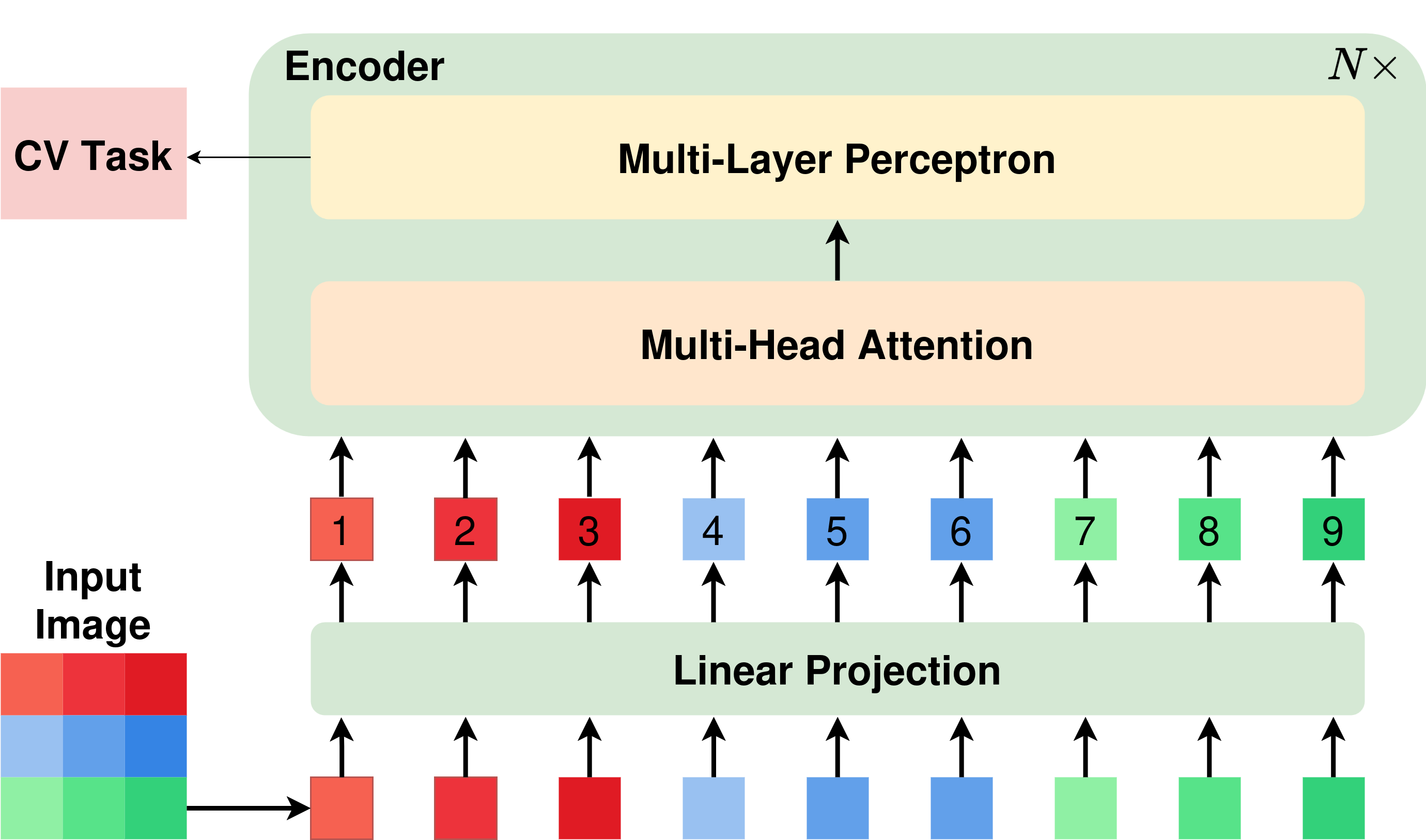}
    \caption{Vision Transformer Architecture.}
    \label{fig:Figure2}
\end{figure}

In ViTs, as shown in Figure~\ref{fig:Figure2}, the transformer architecture is applied directly to visual data, with the input image being divided into fixed-size patches, and each patch being treated as a token with a positional encoding. 
Positional encodings are then added to these tokens to capture the spatial information where the patch is in the image. 
The sequence of tokens then goes through $N$ encoder layers, with the multi-head attention mechanism allowing the model to capture relationships between image patches and the multi-layer perceptron (MLP) further capturing intricate patterns. 
The output of the encoder is then a representation of the input sequence that captures its essential features, which can be used in CV tasks such as image classification and object detection.

A notable evolution of this architecture is the Swin Transformer (Swin-T)~\cite{swin}, which introduces a hierarchical structure and performs self-attention within shifted local windows rather than across the entire image. This design significantly reduces attention complexity while preserving cross-window interaction.

This family of ViT models, which we will refer to as ``Standard ViT" models, quickly became incredibly popular~\cite{han_survey_2023,dong_cswin_2022,arnab_vivit_2021} with their ability to understand global context, often outmatching CNNs in their ability to handle complex visual relationships.
However, Standard ViTs have a large amount of parameters and high computation costs~\cite{xu_devit_2023}, with memory- and computationally- straining calculations being the core of the architecture, primarily because their execution is heavily dominated by parameter-dense FC layers. 
This makes it challenging to run them on resource-constrained edge devices, which would greatly benefit from their strong representational power and state-of-the-art performance in CV tasks~\cite{ImageNet}.

\subsection{Hybrid Vision Transformers}

The Hybrid family of ViT models merges a different set of computational patterns. These models, such as MobileViT~\cite{mobilevit} and EfficientViT~\cite{efficientvit} were designed specifically to maximize efficiency for on-device inference. They achieve this by combining two types of operations: i) convolutions to extract local, low-level features; and ii) lightweight transformer blocks to capture global context. 
As a result, their computational profile is not dominated by one operation type. Instead, they present a balanced and heterogeneous mix of both CONV and various types of FC layers found in transformers.
Therefore, we refer to these architectures as Hybrid ViTs.
Because FC and CONV layers have distinct memory access and data reuse patterns, static accelerators suffer severe underutilization across these diverse workloads, necessitating a flexible hardware design.


\subsection{Hardware Accelerators for Edge ViTs}

The need for flexible, high-performance acceleration is particularly critical in the domain of edge computing~\cite{7488250}. To this end, FPGAs, with their reprogrammable architecture and ability to enhance performance efficiently, are particularly well-suited for accelerating ViT models on power and resource-constrained edge devices. 
On highly resource-constrained devices, such as the PYNQ-Z2 board used in this work, which are often based on low-cost SoCs like the AMD Zynq-7000, the FPGA fabric provides a critical capability for parallel, power-efficient computation that on-chip ARM processors cannot match. Therefore, the critical design challenge is to architect a solution that operates with maximum efficiency within the limited logic and on-chip memory constraints inherent to these edge-FPGAs. 

To design and implement our FlexViT accelerator, we use \textit{SECDA-TFLite}~\cite{Haris2023JPDC}, a toolkit based on the SECDA methodology~\cite{Haris2021SBACPAD} that reduces the design time of specialized FPGA-based accelerators for DNN inference on edge devices.
To achieve this, the \textit{SECDA-TFLite} employs the TensorFlow Lite (TFLite) delegate system, enabling the execution of model components or the entire model on specialized hardware.
This is particularly advantageous as it enables end-to-end evaluation with real model graphs and runtime operator shapes, which is crucial for accurately profiling the irregular dimensions of Hybrid ViTs.

\section{Acceleration of Heterogeneous ViT models}
\label{sec:FlexibleAcceleration}


This section presents the unified hardware–software co-design strategy underlying our flexible FPGA-based accelerator, and describes the key architectural optimizations that enable efficient support for modern hybrid ViT models.


\subsection{Layer Heterogeneity}

The architectural diversity of edge ViT models is quantitatively confirmed by profiling five models using the SECDA-TFLite~\cite{Haris2023JPDC} benchmarking tool, as can be seen in Table~\ref{tab:hybrid_layer_percentages}. 
This data illustrates a stark contrast in computational requirements: Standard ViT models like ViT-T and DeiT-T spend the majority of their runtime in FC layers, with CONV layers being negligible, while Hybrid models like MobileViT-S and EfficientViT-b1 show a much more balanced profile. 

To support all these ViT models, we created a custom TFLite delegate that intercepts FC and CONV operations at runtime. At this stage, we perform an \textit{im2col} (image-to-column) transformation on the CPU which linearizes the 3D convolution input into a 2D matrix format. Consequently, both standard FC layers and CONV layers map to a single, unified GEMM primitive (Input $\times$ Weights $+$ Bias).

While the ``Other" category in Table~\ref{tab:hybrid_layer_percentages} accounts for a significant portion of inference time, particularly in hybrid models, we deliberately exclude these operations from FPGA offloading to preserve generality and maximize resource utilization. 
Implementing diverse non-linear operators such as Softmax and LayerNorm in hardware would require dedicated special function units, consuming valuable logic and DSP resources on the edge FPGA on the PYNQ-Z2 board. We would consider doing such an architectural trade-off if a specific non-linear operation constituted a dominant, consistent bottleneck across all ViT variants (standard and hybrid). 

By restricting the accelerator to a GEMM primitive, we avoid over-specializing the datapath from specific model architectures. This design choice strictly decouples the workload: the CPU manages the irregular, memory-bound activations and control flow, allowing the FPGA resources to be dedicated entirely to maximizing the parallelism of the compute-intensive FC and CONV layers.


\subsection{Dimensional Variance}
 
Standard and hybrid ViT models contain massive variance in layer dimensions. For example, the sequence length in EfficientViT plummets from 12,000 tokens in early stem blocks down to just 49 in later stages, while channel widths across the models range anywhere from 16 in early MobileViT layers up to 1536 in EfficientViT.
After the \textit{im2col} transformation (for CONV layers), each delegated layer is represented as a GEMM with dimensions $(N,M,K)$, where $N$ is the number of input rows (often from the spatial image patches), $M$ is the number of output columns (filters or neurons) and $K$ is the reduction dimension. 
Standard ViT projection (i.e., FC) layers typically yield large, balanced matrices with substantial width $(M)$. Conversely, the CONV layers in hybrid models like MobileViT frequently generate "tall and skinny" matrices where the spatial dimension $N$ is dominant, but the channel width $M$ is comparatively narrow. A static dataflow (e.g., fixed weight-stationary) suffers from severe hardware underutilization and bandwidth bottlenecks when executing across such diverse shapes.

\begin{table}[t!]
\small
\centering
\caption{Total inference time (\%) on the ZYNQ-7000 CPU.}
\label{tab:hybrid_layer_percentages}
\begin{tabular}{|l|l|ccc|}
\hline
\textbf{Type} & \textbf{Model} & \textbf{\makecell[l]{FC\\(\%)}} & \textbf{\makecell[l]{CONV\\(\%)}} & \textbf{\makecell[l]{Other\\(\%)}} \\ \hline
Standard & ViT-T           & 45.14          & 1.22             & 53.64            \\
Standard & DeiT-T          & 45.14          & 1.23             & 53.63            \\
Standard & Swin-T          & 50.45          & 0.33             & 49.22            \\ \hline
Hybrid   & MobileViT-S     & 16.20          & 29.93            & 53.87            \\
Hybrid   & EfficientViT-b1 & 0.54           & 40.29            & 59.17            \\ \hline
\end{tabular}
\end{table}

To address this, our custom TFLite delegate incorporates a host driver that manages a dynamic, dual-mode dataflow. We denote the per-core tile sizes along the row and column dimensions as $T_N$ and $T_M$, respectively, and the number of parallel cores as $C$. The host driver pads the layer dimensions to $(\tilde N, \tilde M, \tilde K)$, selects one of the two execution modes (i.e., Input-Broadcast or Weight-Broadcast) before launching each delegated layer. The hardware scheduler then realizes that choice by changing both the tile traversal order and the operand broadcast pattern to maximize data reuse and bandwidth efficiency.

\textbf{Input-Broadcast Mode} (input broadcast, weight partitioning): It is used when the output width is large enough to keep all cores busy (i.e., $\tilde M \ge C T_M$), or when the Input-Broadcast schedule yields lower estimated DMA traffic. 
In this mode, the outer loop advances along $\tilde N$ in steps of $T_N$, so one input tile is loaded once and broadcast to all cores. The inner loop then advances along $\tilde M$ in steps of $C T_M$, and each core receives a different weight tile. This schedule maximizes activation reuse and is especially effective for FC-dominant layers.

\textbf{Weight-Broadcast Mode} (weight broadcast, input partitioning): It is used when the layer is too narrow to fully utilize the output-width parallelism of the Input-Broadcast mode, or when the Weight-Broadcast schedule yields lower estimated DMA traffic. 
The outer loop advances along $\tilde M$ in steps of $T_M$, so one weight tile is loaded once and broadcast to all cores. The inner loop then advances along $\tilde N$ in steps of $C T_N$, and each core receives a different input tile. This schedule maximizes weight reuse and is more suitable for CONV layers with a large spatial extent.

The logic for this adaptable dataflow is formalized in Algorithm~\ref{alg:dual_mode}. By selecting the optimal dataflow per layer, our accelerator ensures high utilization of the processing elements regardless of whether the model is FC- or CONV-dominated.

\begin{algorithm}[t]
\small
\caption{Dynamic Per-Layer Mode Selection}
\label{alg:dual_mode}
\begin{algorithmic}[1]
\STATE \textbf{Input:} layer type $L \in \{\texttt{FC},\texttt{CONV}\}$
\STATE \textbf{Input:} padded dimensions $(\tilde N,\tilde M,\tilde K)$
\STATE \textbf{Const.:} per-core tile sizes $(T_N,T_M)$ and core count $C$
\STATE \textbf{Const.:} hardware dense width $W_{dense} \gets C \times T_M$

\IF{$L = \texttt{FC}$}
    \STATE \textit{// Spatial heuristic for projection layers}
    \IF{$\tilde M \ge W_{dense}$}
        \STATE $mode \gets \textit{Input-Broadcast}$
    \ELSE
        \STATE $mode \gets \textit{Weight-Broadcast}$
    \ENDIF
\ELSE
    \STATE \textit{// Analytical DMA estimation for high-variance CONV layers}
    \STATE $b_{IB} \gets \textsc{EstimateTransferBytes}(\textit{IB}, \tilde N, \tilde M, \tilde K)$
    \STATE $b_{WB} \gets \textsc{EstimateTransferBytes}(\textit{WB}, \tilde N, \tilde M, \tilde K)$
    \IF{$b_{IB} \le b_{WB}$}
        \STATE $mode \gets \textit{Input-Broadcast}$
    \ELSE
        \STATE $mode \gets \textit{Weight-Broadcast}$
    \ENDIF
\ENDIF

\STATE \textbf{Return} $mode$
\end{algorithmic}
\end{algorithm}


\subsection{Resource Limitations}

To sustain single-pass accumulation on a resource-constrained edge FPGA, FlexViT fixes the execution granularity of each core and sizes the on-chip buffers accordingly. We denote the per-core tile sizes by $T_N$ and $T_M$, the maximum buffered reduction depth by $T_K$, and the single-instruction multiple-data (SIMD) width inside each core by $K_f$.
This implements single-pass accumulation where the input and weight buffers are sized to hold the full depth of a tile, allowing the PE array to accumulate the full dot-product in local registers.
Because this accumulation is completely local, we never write 32-bit partial sums to main memory; only the final, quantized 8-bit values are written out.

To support this strategy within the specific constraints of the PYNQ-Z2 board, we design the hardware accelerator with the following parameters, selected to maximize throughput while fitting within the resources:

\begin{itemize}    
    \item \textbf{Per-core tile size} ($T_N = 64$, $T_M = 64$): each core computes a $64 \times 64$ output tile. With $C = 3$ cores, the Input-Broadcast mode covers up to $3T_M = 192$ output columns per inner-loop step, while the Weight-Broadcast mode covers up to $3T_N = 192$ input rows per inner-loop step. This granularity is large enough to expose useful parallelism on FC-heavy layers while keeping loop-control overhead manageable for CONV layers.
    
    \item \textbf{Buffer Depth $(T_K = 1024)$:} Crucially, we size the on-chip input and weight buffers to store a depth of $T_K$=1024. This capacity covers $100\%$ of the layers in ViT, DeiT, Swin and MobileViT models. In EfficientVIT, it accommodates the entire feature extraction backbone (44 layers), excluding only the single final classification layer (with $T_K=1536$). This ensures that for four out of the five evaluated ViT models, the accelerator executes without any depth-wise tiling, completely eliminating partial-sum swapping.
        
    \item \textbf{Parallelism (C=3):} We instantiate 3 parallel compute cores to maximize compute density on the PYNQ-Z2 board. While scaling to 4 cores would exceed the device's BRAM capacity, 3 cores optimally utilize the available on-chip memory. To support this configuration, the local buffer width is sized to maximize the BRAM bandwidth, ensuring simultaneous, high-throughput access to weights and inputs without stalling the compute engines.
\end{itemize}

\section{FlexViT Accelerator}
\label{sec:flexViT}

We now describe the FlexViT hardware accelerator shown in Figure~\ref{fig:secda-vit_arch}. FlexViT is implemented with six customized hardware units: the scheduler; the read units (ReadInp for inputs, ReadWgt for weights, and ReadBias for the bias and requantization parameters); the high-throughput GEMM engine; and the post-processing unit for processing, requantizing and sending data back to the host driver. 
Table~\ref{tab:flexvit_params} summarizes the architectural parameters used throughout this section. 

\begin{table}[t]
\centering
\caption{FlexViT architectural parameters.}
\label{tab:flexvit_params}
\footnotesize
\begin{tabular}{|c|c|c|}
\hline
\textbf{Symbol} & \textbf{Value} & \textbf{Description} \\
\hline
$T_N$ & 64 & Per-core row tile size \\
\hline
$T_M$ & 64 & Per-core output-channel tile size \\
\hline
$T_K$ & 1024 & Maximum buffered reduction depth \\
\hline
$C$ & 3 & Number of parallel GEMM cores \\
\hline
$K_f$ & 16 & SIMD width along the reduction dimension \\
\hline
\end{tabular}
\end{table}




\subsection{Scheduler}

\begin{figure}[t]
    \centering    
    \begin{subfigure}[c]{0.35\textwidth}
        \centering
        \includegraphics[width=\linewidth]{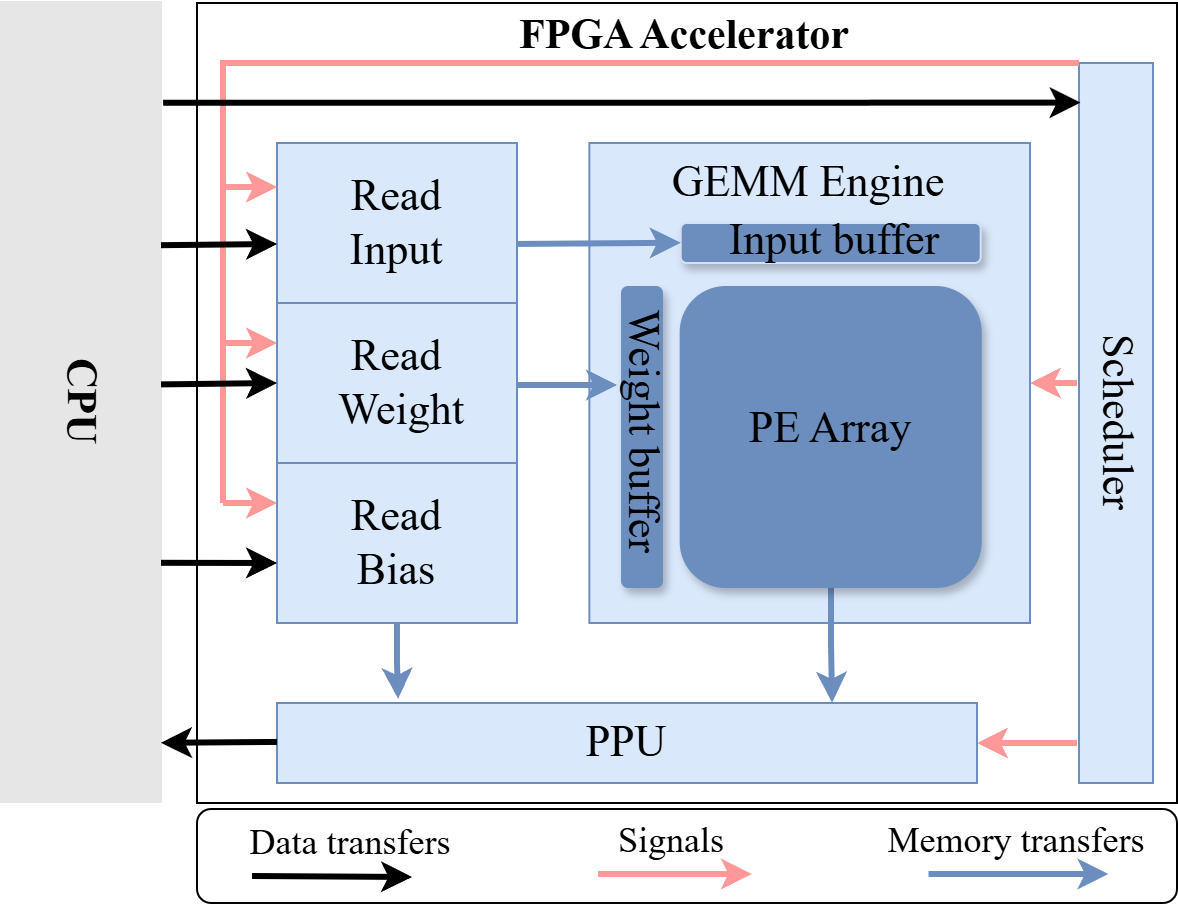}
        \caption{FlexViT Hardware Accelerator.}
        \label{fig:secda-vit_arch}
    \end{subfigure}
    \hfill
    \begin{subfigure}[c]{0.45\textwidth}
        \centering
        \includegraphics[width=\linewidth]{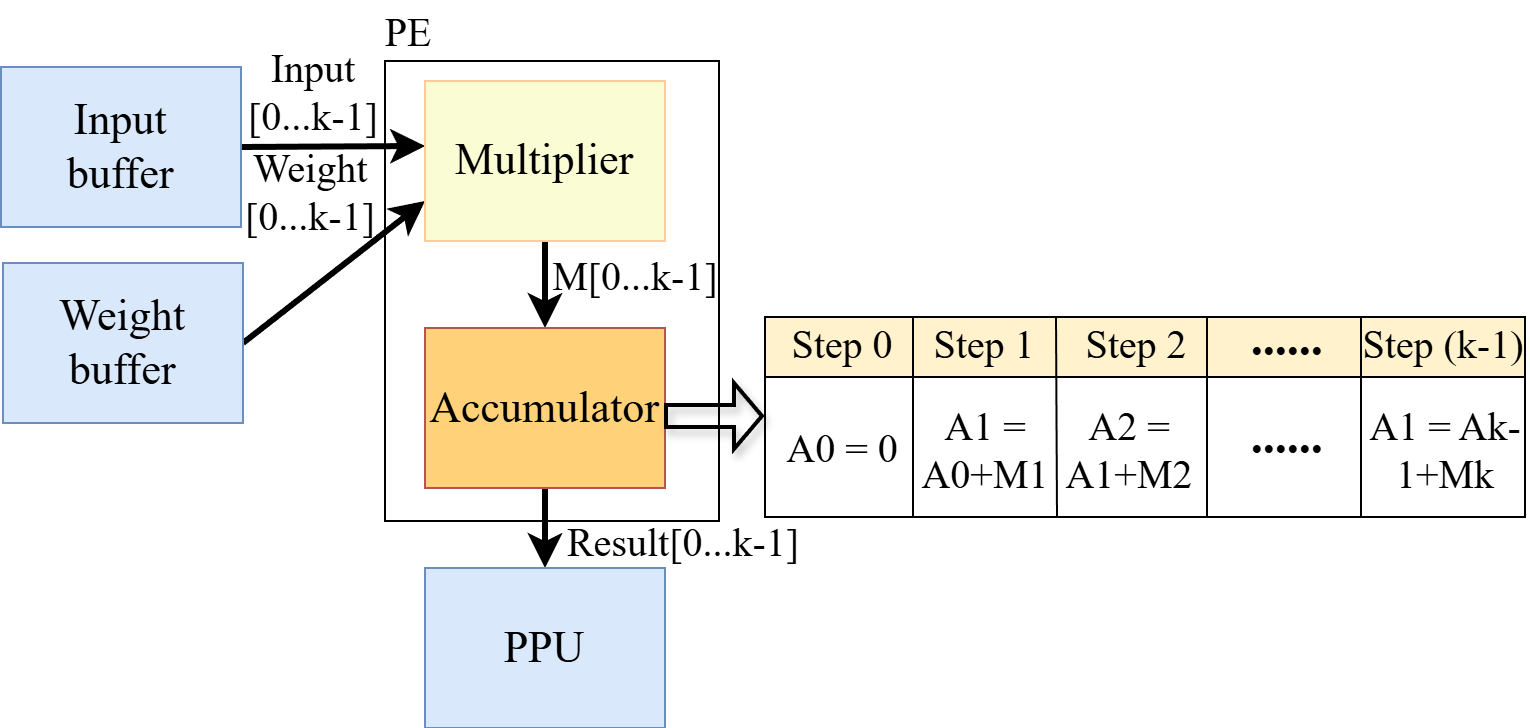}
        \caption{Computational flow within the GEMM Engine.}
        \label{fig:secda-vit_ComFlow}
    \end{subfigure}
    \caption{FlexViT architecture and computational flow.}
    \label{fig:flexvit_overview}
\end{figure}

The operation of the accelerator is controlled by a central Scheduler unit (``Scheduler" in Figure~\ref{fig:secda-vit_arch}), which orchestrates all the accelerator units. Unlike static accelerators that follow a fixed dataflow, the FlexViT scheduler implements dynamic control logic to support both Input-Broadcast and Weight-Broadcast dataflows. Execution begins by reading a layer configuration packet from a dedicated AXI-stream channel containing the layer dimensions, mode flags and quantization parameters. 

Control flow is enforced via a decoupled handshake protocol. The scheduler asserts dedicated validity signals to trigger specific pipeline stages. Each stage operates autonomously and acknowledges completion only when its local task is finished. This decentralized synchronization allows the scheduler to dynamically mask the variable latency of DRAM access, preventing the entire pipeline from stalling due to a single delayed memory transaction. 

Once configured, the scheduler operates at tile granularity exactly as described in Algorithm~\ref{alg:dual_mode}. In the Input-Broadcast mode, it iterates over $\tilde N$-tiles in the outer loop and over $\tilde M$-tiles in the inner loop; in the Weight-Broadcast mode, this loop order is reversed. The arithmetic datapath is unchanged across the two modes. Instead, the scheduler changes whether an incoming tile is broadcast to all $C$ cores or partitioned across them.

The scheduler triggers the input, weight and bias reading units concurrently, using separate AXI streams for each resource and thus allowing the units to operate without read contention.
The PPU processes the results of the current tile \textit{t} while the read units simultaneously fetch the data for tile \textit{t+1}; this overlapping minimizes the latency of data movement, ensuring the compute cores remain saturated.


\subsection{Read Units}

Prior to inference, FlexViT employs a data preloading mechanism that stages all model weights from off-chip DRAM directly into DMA buffers. By decoupling this bulk memory transfer from the active execution phase, FlexViT masks the initialization latency and minimizes the runtime overhead of copying data. 
To maximize throughput, the accelerator is designed to exploit task-level parallelism by managing data dependencies. It achieves this by first using four different AXI streams, each for its own resource: one stream for layer metadata, one for inputs, one for weights, and one for bias data. This allows us to maximize overlap by allowing inputs, weights, and biases to be read concurrently. To efficiently read contiguous data blocks from main memory, the read units are configured to issue AXI4 burst transactions of 32 bits.



The ReadInp and ReadWgt units are responsible for fetching data from main memory to populate the accelerator's on-chip and weight buffers. To support the distinct scheduling strategies defined in~\ref{alg:dual_mode}, which dictate whether inputs or weights are reused, these units dynamically adjust how they fill and distribute data from these buffers to the compute array.
The two units are similar in design and behavior. 
In Input-Broadcast mode, ReadInp fills a shared buffer where one input tile is replicated across all $C$ cores, while ReadWgt populates partitioned buffers so that each core receives a different weight tile. 
In Weight-Broadcast mode, ReadWgt fills a shared buffer where one weight tile is replicated across all $C$ cores, while ReadInp populates partitioned buffers so that each core receives a different input tile. Therefore, the difference between the two modes lies in how the on-chip buffers are managed and routed, rather than in the compute core itself.
Simultaneously, the ReadBias thread reads the bias if it is available for a layer. 
Subsequently, it also reads and processes the necessary requantization variables, such as the quantization offsets (zero-point correction terms) for the layer, which will be required for the PPU.

To support the concurrent access of the multi-core architecture, the on-chip memory hierarchy is banked. We utilize a cyclic partitioning strategy for both the input and weight buffers, interleaving consecutive data elements across separate physical memory banks in a round-robin fashion. This alignment ensures that the vector processing units can fetch contiguous data blocks in a single clock cycle, effectively eliminating the bank conflicts that could otherwise stall the pipeline during high-bandwidth parallel access.



\subsection{GEMM Engine}

The computational core of FlexViT is the GEMM engine, instantiated as three parallel cores. Each core is a vector processing unit designed to execute a tiled dot-product over the common dimension, $K$, using an output-stationary dataflow, where intermediate partial sums remain fixed within the processing elements while inputs and weights are streamed through, eliminating partial-sum traffic to on-chip memory. 
To minimize the limited memory port contention, intermediate accumulators are maintained in local registers rather than BRAM. The datapath executes a continuous, pipelined accumulation along the $K$ dimension, retiring MAC operations on every clock cycle to resolve the full dot-product within the local register files.

To optimize resource usage on the PYNQ-Z2, the PE unit employs a hybrid arithmetic split architecture. The inner compute loop is structurally divided: the 8 bits of the dot-product are synthesized onto the FPGA's logic fabric (LUTs), while the upper 8 bits map to the hardened DSP slices. 
This heterogeneity allows FlexViT to increase its effective MAC density per clock cycle. 
Consequently, FlexViT increases its effective local MAC density, retiring more MAC operations per clock cycle than a standard DSP-only implementation could on the same resource-constrained SoC.
Internally, all partial sums are accumulated into 32-bit wide registers, ensuring that precision is maintained across the depth of the dot-product before the final quantization step.

\noindent\textbf{Computational Flow}: 
As shown in Figure~\ref{fig:secda-vit_ComFlow}, computation is distributed across parallel PEs, with each core operating as a SIMD vector engine whose inner loop is unrolled along the $K$ dimension by the SIMD width $K_f=16$. 
At each cycle, a PE reads $K_f$ activations and the corresponding weight vector from the banked buffers and performs a fully pipelined INT8 MAC, with a latency of 7 cycles and an initiation interval of 1 cycle. Repeating this process for $T_K$ iterations completes the tile-depth traversal while keeping all intermediate partial sums in local registers, thereby realizing an output-stationary dataflow that avoids partial-sum traffic to on-chip memory. Because the software-side \textit{im2col} transformation already normalizes heterogeneous layers into GEMMs, Input-Broadcast and Weight-Broadcast modes differ only in operand distribution and partitioning, rather than in the arithmetic datapath.

\subsection{Post-Processing Unit (PPU)}

The PPU is the final component that serves as a convergence point for the accelerator's parallel architecture.
To maintain throughput while processing requantization, the PPU is deeply pipelined with a latency of 29 cycles and an initiation interval of 2 cycles, due to the several contending read/write operations. 
A key architectural optimization of this design is decoupling the bias and quantization offset (i.e., a quantization parameter) calculations from the main compute path. A naive, sequential datapath would create a long data dependency, forcing the compute engine to stall while bias and quantization offset data are fetched from memory. Our design overcomes this by splitting the operation into two concurrent datapaths: one for the PE and one for the Bias, which are executed in parallel.

Within the PPU, we handle quantization for both CONV and FC layers.
This can be dynamically selected based on the \textit{layer} control signal from the scheduler, and this selection is the core of our ``flexible" accelerator. 
The quantization parameters that were read through the ReadBias unit are used in this step. 
If the layer is \textbf{FC}, the value is routed to a per-tensor requantization block, which uses the single scaling factor required for FC layers in TFLite. 
If the layer is \textbf{CONV}, the value is routed to a per-channel requantization block, which accesses an array of scaling factors to handle the quantization parameters required by CONV layers in TFLite.
Finally, the output from this flexible quantization gets packed into 32-bit packets, written back through the AXI-Stream and sent back to the host.


Finally, to scale the 32-bit accumulated outputs from the GEMM engine down to the required 8-bit precision, the PPU performs fixed-point rescaling. This is efficiently executed using a 32-bit integer multiplication mapped to standard DSP slices, followed by a bitwise right-shift $(S\approx M \cdot 2^{-n})$.

\section{Evaluation}
\label{sec:evaluation}

\subsection{Experimental Setup}
\label{sec:evaluation-experimental_setup}

To design and evaluate our FlexViT accelerator, we utilized the SECDA-TFLite toolkit~\cite{Haris2023JPDC} and the PYNQ-Z2 board, which contains an AMD Zynq-7000 edge SoC with an edge FPGA and an ARM Cortex-A9 CPU~\cite{pynqz2}. 
For performance and power comparisons, we evaluate against the on-board CPU with NEON SIMD instructions enabled, over an average of 100 runs. The design is synthesized with a clock frequency of 200MHz. 

To ensure a fair comparison and avoid potential static power overhead from the FPGA fabric, CPU-only power measurements were were collected without programming the FlexViT hardware logic onto the FPGA. Instead, the FPGA was configured with a minimal baseline bitstream containing only the system design required to benchmark the CPU. Consequently, the static power overhead introduced by the FPGA fabric during CPU-only execution remains negligible.

FlexViT was evaluated using five ViT models converted to fully quantized INT8 TFLite format. ViT-T~\cite{vit-tiny} is pretrained on the ImageNet-21k dataset~\cite{ImageNet}, whereas DeiT-T~\cite{DeiT}, Swin-T~\cite{swin}, MobileViT-S~\cite{mobilevit}, and EfficientViT-b1~\cite{efficientvit} are pretrained on the ImageNet-1k dataset~\cite{ImageNet}. The parameter counts and model sizes of the five models are summarized in Table~\ref{tab:model_parameters}. 
The evaluated input resolutions ($224 \times 224$ and $256 \times 256$) are highly practical for edge applications such as industrial IoT systems like automated defect detection or smart-home platforms.

Note that FlexViT executes unmodified INT8 TFLite models and therefore preserves the inference accuracy of the software baseline. To validate functional correctness, we verify that the inference outputs produced by the hardware accelerator for all five models match those from CPU execution within TFLite's acceptable tolerance (cosine similarity $>$ 99\%). Therefore, we evaluate the proposed accelerator using three metrics: resource utilization (\%), end-to-end latency (s), and average energy per inference (J).

\begin{table}[t]
\small
\centering
\caption{ViT models under study.}
\label{tab:model_parameters}
\begin{tabular}{|c|c|c|c|}
\hline
\textbf{\makecell[c]{Model}} & \textbf{\makecell[c]{Params\\(M)}} & \textbf{\makecell[c]{Size\\(MB)}} & \textbf{\makecell[c]{Image\\Size}} \\ \hline

ViT-T            & 5.7 & 5.7 & $224\times224$ \\
DeiT-T           & 5.7 & 5.7 & $224\times224$ \\
Swin-T           & 28     &  27.7    &  $224\times224$      \\
MobileViT-S      & 5.6 & 5.7 &  $256\times256$ \\
EfficientViT-b1 & 9.1 & 9.6 &  $224\times224$ \\ \hline
\end{tabular}
\end{table}


\subsection{Results}


\subsubsection{Latency and Energy}

The latency results are presented in Table~\ref{tab:results_table}. As we can see, FlexViT accelerates the offloaded layers by up to $2.74\times$. 
By efficiently offloading the most compute-intensive FC and CONV layers, the hardware drastically reduces the inference bottleneck across a diverse spectrum of ViT families. This efficient local execution translates to consistent and verifiable speedups of up to $1.40\times$ across evaluated models.

The variation in end-to-end speedups is a direct reflection of the underlying model architecture rather than a limitation in the accelerator's throughput. Specifically, the speedups align closely with the computational profiles established in our layer heterogeneity analysis (Table~\ref{tab:hybrid_layer_percentages}). For standard ViT models like ViT-T and DeiT-T, where over 45\% of the baseline execution is dominated by dense FC layers, the accelerator can sustain long periods of uninterrupted execution. This results in significant layer-wise speedups ($2.61\times$ for ViT-T and $2.74\times$ for DeiT-T), driving the highest end-to-end gains ($1.40\times$).

Crucially, the accelerator proves equally adept at handling the complex, heterogeneous layers of hybrid ViT models. At the layer level, FlexViT achieves speedups of $1.77\times$ and $2.00\times$ for MobileViT-S and EfficientViT-b1, respectively. 
However, these strong layer-wise gains translate into more modest end-to-end speedups of $1.25\times$ and $1.24\times$. Such system-level improvements are naturally bounded by Amdahl's law~\cite{amdahl}, since more than 50\% of the original runtime of these models is spent on non-offloaded operations (such as SoftMax, LayerNorm and memory reshaping) that remain strictly executed on the CPU.
We note that the execution time of these non-offloaded ``Other" operations remains practically identical across baseline and accelerated runs; the marginal differences fall well within standard measurement variance across our 100 runs and reflect the negligible software-side overhead of the runtime driver transitioning.
This demonstrates that as hybrid ViT models fragment their execution graphs with diverse scalar operations, the CPU's sequential execution becomes the overriding system bottleneck, regardless of the accelerator's efficiency.

\begin{table*}[t]
\centering
\fontsize{5.5}{7.5}\selectfont
\caption{Latency and energy results on CPU and CPU + Accelerator (100 runs).}
\label{tab:results_table}
\resizebox{1.0\linewidth}{!}{
\begin{tabular}{|l|l|c|c|c|c|c|c|c|c|}
\hline
\textbf{Model} & \textbf{HW} & \textbf{FC (s)} & \textbf{CONV (s)} & \textbf{Other (s)} & \textbf{Total (s)} & \textbf{\shortstack{Acc.\\Speedup}} & \textbf{\shortstack{E2E\\Speedup}} & \textbf{Energy (J/inference)} & \textbf{\shortstack{Energy\\Gain}} \\ 
\hline
\multirow{2}{*}{\textbf{ViT-T}} & CPU & 0.92 & 0.02 & 1.10 & 2.04 & \multirow{2}{*}{\textbf{2.61$\times$}} & \multirow{2}{*}{\textbf{1.40$\times$}} & 3.53 & \multirow{2}{*}{\textbf{1.07$\times$}} \\
 & CPU+Acc & 0.35 & 0.01 & 1.10 & 1.46 & & & 3.31 & \\ \hline
\multirow{2}{*}{\textbf{DeiT-T}} & CPU & 0.93 & 0.03 & 1.08 & 2.04 & \multirow{2}{*}{\textbf{2.74$\times$}} & \multirow{2}{*}{\textbf{1.40$\times$}} & 3.56 & \multirow{2}{*}{\textbf{1.09$\times$}} \\
 & CPU+Acc & 0.34 & 0.01 & 1.11 & 1.46 & & & 3.28 & \\ \hline
\multirow{2}{*}{\textbf{Swin-T}} & CPU & 3.87 & 0.03 & 3.77 & 7.67 & \multirow{2}{*}{\textbf{2.05$\times$}} & \multirow{2}{*}{\textbf{1.35$\times$}} & 12.96 & \multirow{2}{*}{\textbf{1.03$\times$}} \\
 & CPU+Acc & 1.88 & 0.02 & 3.78 & 5.68 & & & 12.56 & \\ \hline
\multirow{2}{*}{\textbf{\shortstack{Mobile\\ViT-S}}} & CPU & 0.69 & 1.28 & 2.30 & 4.27 & \multirow{2}{*}{\textbf{1.77$\times$}} & \multirow{2}{*}{\textbf{1.25$\times$}} & 7.31 & \multirow{2}{*}{\textbf{0.98$\times$}} \\
 & CPU+Acc & 0.27 & 0.84 & 2.31 & 3.42 & & & 7.49 & \\ \hline
\multirow{2}{*}{\textbf{\shortstack{Efficient\\ViT-b1}}} & CPU & 0.01 & 0.69 & 1.01 & 1.71 & \multirow{2}{*}{\textbf{2.00$\times$}} & \multirow{2}{*}{\textbf{1.24$\times$}} & 2.95 & \multirow{2}{*}{\textbf{0.93$\times$}} \\
 & CPU+Acc & 0.01 & 0.34 & 1.027 & 1.377 & & & 3.16 & \\ \hline
\end{tabular}
}
\end{table*}

Energy measurements can also be seen in Table~\ref{tab:results_table}. We gather energy metrics by using a Makerfire USB power meter~\cite{makerfile}. 
Again, the energy efficiency of our system is closely tied to the percentage of layers offloaded to the accelerator. Encouragingly, the design delivers consistent end-to-end latency improvements while keeping the total energy per inference nearly identical to, or better than, the CPU baseline.
The energy outcomes highlight the trade-offs of hardware-software co-design on edge SoCs. For compute-intensive models like DeiT-T, the significant reduction in overall execution time directly yields net-positive energy savings ($1.09\times$). 

In the worst-case scenarios (such as EfficientViT-b1), where $\sim$41\% of the runtime is offloaded to the FPGA, the dynamic power of the FPGA fabric results in a slight increase in average energy per inference compared with the CPU-only baseline ($0.93\times$). This minimal energy overhead represents a trade-off for achieving a $1.24\times$ speedup.

This energy penalty in hybrid architectures arises because both the CPU and FPGA fabric remain active throughout inference: the FPGA sustains baseline power during CPU-dominated phases, whereas the CPU remains active during FPGA execution. This highlights the power overhead of heterogeneous workloads with limited FPGA utilization.

\subsubsection{Resource Utilization(\%)}

\begin{figure}[t]
    \centering
    \includegraphics[width=0.95\linewidth]{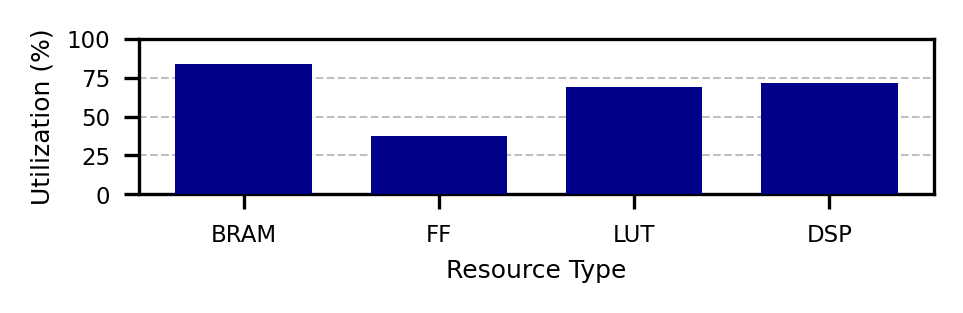}
    \caption{FlexViT resource utilization on PYNQ-Z2 board.}
    \label{fig:resource-util}
\end{figure}

Figure~\ref{fig:resource-util} shows the resource utilization of FlexViT on the FPGA fabric of the PYNQ-Z2 board. The proposed design consumes $83.9\%$ of the available BRAM, $71.8\%$ of the DSPs, and $69.3\%$ of the LUTs, indicating a dense and efficient mapping to the target device. 
This resource utilization reflects a deliberate design strategy to maximize computational throughput within the strict logic and memory constraints of an edge-class FPGA, while maintaining a balanced allocation across the available resources. The role of each resource is discussed below:

\begin{itemize}
    \item \textbf{BRAM:} The near-full BRAM utilization (83.9\%) confirms that our tiling strategy maximizes the on-chip tile size, enabling single-pass accumulation. By capturing the largest possible working set on-chip, the design actively minimizes the latency overhead associated with off-chip memory transactions.
    
    \item \textbf{DSP:} The high DSP utilization (71.8\%) reflects the dense parallelism of the GEMM core. This deliberate mapping confirms that the architecture is optimized to extract the maximum available computational throughput from the limited resources of an edge platform.
    
    \item \textbf{LUT:} The substantial LUT utilization (69.3\%) is primarily driven by two factors: the logic required for the dual-mode scheduler to dynamically route data between cores, and the split architecture, which offloads a portion of the arithmetic operations to LUTs, increasing the effective MAC throughput beyond the physical DSP limit. This illustrates a balanced mapping strategy that leverages general logic to supplement scarce arithmetic resources without causing routing congestion.
\end{itemize}

\subsection{Comparison with State-of-the-Art}
\label{sec:comparison_sota}

While prior ViT accelerators~\cite{10557992,m2vit,via} have demonstrated performance on powerful FPGA platforms, such as the ZCU102 and Alveo U50, these evaluations typically assume higher resource availability and power budgets, with reported power consumption ranging from 18.2 to 39 W.
This leaves a gap in understanding the practicality of ViT acceleration under resource-constrained edge devices. 
FlexViT addresses this deployment scenario by targeting measured, end-to-end inference on a resource-constrained edge platform, the PYNQ-Z2 board.
FlexViT achieves execution at only 2.19 to 2.29 W while reporting end-to-end CPU+FPGA system-level performance.
Furthermore, through a unified FC/CONV GEMM-based accelerator architecture, FlexViT supports both standard and hybrid ViT models, demonstrating its suitability for practical edge AI deployment. 
Because raw quantitative metrics alone may not capture these practical deployment considerations, we compare FlexViT with state-of-the-art works along three critical system-level dimensions: (1) end-to-end verifiability, (2) resource efficiency under edge constraints, and (3) architectural coverage.



\noindent\textbf{End-to-end verifiability vs. simulation:} A key distinction of FlexViT is that all reported performance results are measured on a deployed PYNQ-Z2 platform, inherently capturing system-level overheads such as DRAM latency, AXI bus contention, and CPU-FPGA synchronization. 
In contrast, works targeting similar hardware, such as ViTA~\cite{vita}, report throughput estimates derived from synthesis results rather than physical execution. Their methodology relies on estimated power consumption and analytically derived frame rates based on idealized datapath behavior. Consequently, these evaluations do not capture system-level effects encountered during real deployment. 
Furthermore, works such as Cao \textit{et al.}~\cite{an_energy_efficient} omit key implementation details, including memory mapping and driver interfaces, limiting the reproducibility of their results. By leveraging the SECDA-TFLite framework~\cite{Haris2023JPDC}, FlexViT provides a transparent, fully deployable, and reproducible evaluation methodology based on measured end-to-end performance.

\noindent\textbf{Resource efficiency vs. hardware scaling:} Prior accelerators targeting hybrid ViT models frequently rely on high-end FPGA platforms that are impractical for cost-sensitive edge deployment. For example, Shao \textit{et al.}~\cite{10557992} and M\textsuperscript{2}ViT~\cite{m2vit} target the Xilinx ZCU102, which provides more than $10~\times$ the logic resources of the PYNQ-Z2. 
Although Shao \textit{et al.}~\cite{10557992} report a DSP utilization of only $40.6\%$, this corresponds to more than 1,000 DSP slices on the ZCU102; nearly $5~\times$ the total DSP capacity of our target platform. Moreover, these devices accommodate substantially larger on-chip buffers, reducing off-chip memory traffic and alleviating the bandwidth bottlenecks that characterize resource-constrained edge platforms. Consequently, direct comparisons based solely on conventional efficiency metrics can be misleading across such disparate hardware targets.

\noindent\textbf{Architectural coverage:} Existing accelerators are often tailored to specific ViT model families. ViTA~\cite{vita} is tightly coupled to the matrix-vector multiplication (MVM) operations of attention blocks, preventing support for CONV layers found in modern hybrid ViTs. Conversely, M\textsuperscript{2}ViT~\cite{m2vit} is highly specialized for EfficientViT and therefore offers limited applicability to other architectures. In contrast, FlexViT adopts a unified execution model that maps both FC and CONV layers onto a single GEMM primitive, enabling efficient support for both standard and hybrid ViT models requiring separate compute engines.

\noindent To the best of our knowledge, FlexViT is the first end-to-end FPGA accelerator demonstrated on an edge-class platform to provide verified performance improvements across both standard and hybrid ViT families within a unified architecture.

\section{Related Work}
\label{sec:related_work}

\noindent\textbf{Compression-based Accelerators:} 
FPGA accelerators aim to improve computational throughput and energy efficiency by leveraging algorithmic techniques to minimize model size.
Auto-ViT-Acc~\cite{autovitacc} applies fixed-point and power-of-two quantization to the compute-intensive linear (FC) layers of ViT models to enable efficient acceleration.
VAQF~\cite{vaqf} is a framework that automatically generates FPGA inference accelerators for ViT models with binary weights and low-precision activations, significantly reducing memory requirements for model parameters and intermediate data. 
Several ViT works also employ token pruning~\cite{heatvit, staticanddynamicpruning, vitcod} and token merging~\cite{tome, adaptiv} techniques to reduce input sequence length and ViT model complexity, often improving throughput and energy efficiency. 
DRViT~\cite{drvit} combines token merging, token pruning, and INT8 quantization with a specialized hardware accelerator. 

While these approaches demonstrate impressive efficiency improvements, they typically require substantial model modifications and additional hardware specialization. 
In contrast, FlexViT avoids such modifications by relying on standard INT8 quantization, enabling the acceleration of off-the-shelf ViT models without custom retraining or complex pruning.

\noindent\textbf{Dataflow-based Accelerators:} 
Prior FPGA accelerators for ViT models have primarily explored dataflow optimization and operator specialization~\cite{lightweight, via, outputblock, 109, mevit}. However, many designs target narrow workload patterns, such as FC-dominant Standard ViTs, Swin-specific execution~\cite{swat, swin}, or individual hybrid architectures implemented on larger FPGA platforms. 
In contrast, FlexViT enables unified and practical edge deployment by mapping both FC and CONV layers onto a single GEMM engine and validating the complete system end-to-end on a resource-constrained PYNQ-Z2 platform.

\section{Conclusion}
\label{sec:conclusion}

We presented FlexViT, a lightweight and flexible FPGA accelerator for Vision Transformer inference on resource-constrained edge devices. 
FlexViT employs a hardware-software co-design approach that unifies FC and CONV layers found in modern ViT models into a single GEMM primitive, enabling broad support for standard and hybrid ViT architectures. 
By integrating INT8 quantization, hardware optimizations, and a flexible post-processing pipeline, FlexViT achieves up to $2.74\times$ layer-level acceleration and up to $1.40\times$ end-to-end speedup across a range of ViT models deployed on a Zynq-7000 SoC. 
As future work, we plan to explore deeper system-level optimizations to alleviate memory bottlenecks in the current layer-by-layer execution model.




\section*{Acknowledgments}



This work was partially supported by the EU Project dAIEDGE (GA Nr 101120726) and the Innovate UK Horizon Europe Guarantee (GA Nr 10090788).


\balance

\bibliographystyle{IEEEtran}
\bibliography{0_main}

@ARTICLE{7488250,
  author={Shi, Weisong and Cao, Jie and Zhang, Quan and Li, Youhuizi and Xu, Lanyu},
  journal={IEEE Internet of Things Journal}, 
  title={Edge Computing: Vision and Challenges}, 
  year={2016},
  volume={3},
  number={5},
  pages={637-646},
  doi={10.1109/JIOT.2016.2579198}}

@inproceedings{Haris2021SBACPAD,
    title={{SECDA: Efficient Hardware/Software Co-Design of FPGA-based DNN Accelerators for Edge Inference}},
    author={Jude Haris and Perry Gibson and José Cano and Nicolas Bohm Agostini and David Kaeli},
    booktitle = {SBAC-PAD},  
    year={2021},
    pages = {1--8}
}

@article{Haris2023JPDC,
    title = {{SECDA-TFLite}: A toolkit for efficient development of {FPGA}-based {DNN} accelerators for edge inference},
    author = {Jude Haris and Perry Gibson and José Cano and Nicolas {Bohm Agostini} and David Kaeli},
    journal = {Journal of Parallel and Distributed Computing},
    volume = {173},
    pages = {140-151},
    year = {2023},
    issn = {0743-7315},
    doi = {10.1016/j.jpdc.2022.11.005},
}

@inproceedings{vaswani_attention_2023,
author = {Vaswani, Ashish and Shazeer, Noam and Parmar, Niki and Uszkoreit, Jakob and Jones, Llion and Gomez, Aidan N. and Kaiser, \L{}ukasz and Polosukhin, Illia},
title = {Attention is all you need},
booktitle = {Advances in Neural Information Processing Systems},
year = {2017},
isbn = {9781510860964},
publisher = {Curran Associates Inc.},
pages = {6000–6010},
numpages = {11},
location = {Long Beach, California, USA},
series = {NIPS'17}
}

@misc{dosovitskiy_image_2021,
    title = {An {Image} is {Worth} 16x16 {Words}: {Transformers} for {Image} {Recognition} at {Scale}},  
    url = {http://arxiv.org/abs/2010.11929},  
    urldate = {2024-01-09},
    publisher = {arXiv},
    author = {Dosovitskiy, Alexey and Beyer, Lucas and Kolesnikov, Alexander and Weissenborn, Dirk and Zhai, Xiaohua and Unterthiner, Thomas and Dehghani, Mostafa and Minderer, Matthias and Heigold, Georg and Gelly, Sylvain and Uszkoreit, Jakob and Houlsby, Neil},
    month = jun,
    year = {2021},
    note = {arXiv:2010.11929 [cs]},      
}

@article{han_survey_2023,
    title = {A {Survey} on {Vision} {Transformer}},
    volume = {45},
    issn = {1939-3539},
    doi = {10.1109/TPAMI.2022.3152247},    
    number = {1},
    urldate = {2023-12-13},
    journal = {IEEE Transactions on Pattern Analysis and Machine Intelligence},
    author = {Han, Kai and Wang, Yunhe and Chen, Hanting and Chen, Xinghao and Guo, Jianyuan and Liu, Zhenhua and Tang, Yehui and Xiao, An and Xu, Chunjing and Xu, Yixing and Yang, Zhaohui and Zhang, Yiman and Tao, Dacheng},
    month = jan,
    year = {2023},
    pages = {87--110},    
}

@inproceedings{dong_cswin_2022,
    title = {{CSWin} {Transformer}: {A} {General} {Vision} {Transformer} {Backbone} {With} {Cross}-{Shaped} {Windows}},    
    booktitle = {Proceedings of the {IEEE/CVF} {Conference} on {Computer} {Vision} and {Pattern} {Recognition}},
    urldate = {2023-12-13},
    author = {Dong, Xiaoyi and Bao, Jianmin and Chen, Dongdong and Zhang, Weiming and Yu, Nenghai and Yuan, Lu and Chen, Dong and Guo, Baining},
    year = {2022},
    pages = {12124--12134}
}

@inproceedings{arnab_vivit_2021,
    title = {{ViViT}: {A} {Video} {Vision} {Transformer}},    
    booktitle = {Proceedings of the {IEEE/CVF} {International} {Conference} on {Computer} {Vision}},
    urldate = {2023-12-13},
    author = {Arnab, Anurag and Dehghani, Mostafa and Heigold, Georg and Sun, Chen and Lučić, Mario and Schmid, Cordelia},
    year = {2021},
    pages = {6836--6846}
}

@article{xu_devit_2023,
author = {Xu, Guanyu and Hao, Zhiwei and Luo, Yong and Hu, Han and An, Jianping and Mao, Shiwen},
title = {DeViT: Decomposing Vision Transformers for Collaborative Inference in Edge Devices},
year = {2024},
issue_date = {May 2024},
publisher = {IEEE Educational Activities Department},
address = {USA},
volume = {23},
number = {5},
issn = {1536-1233},
doi = {10.1109/TMC.2023.3315138},
journal = {IEEE Transactions on Mobile Computing},
month = may,
pages = {5917–5932},
numpages = {16}
}

@inproceedings{mobilevit,
title={MobileViT: Light-weight, General-purpose, and Mobile-friendly Vision Transformer},
author={Sachin Mehta and Mohammad Rastegari},
booktitle={International Conference on Learning Representations},
year={2022},
}

@manual{pynqz2,
  title        = {{PYNQ-Z2 User Manual v1.0}},
  author       = {{TUL Corporation}},
  year         = {2018},
  month        = {May},
  organization = {TUL Corporation},

}

@inproceedings{
tome,
title={Token Merging: Your ViT But Faster},
author={Daniel Bolya and Cheng-Yang Fu and Xiaoliang Dai and Peizhao Zhang and Christoph Feichtenhofer and Judy Hoffman},
booktitle={The Eleventh International Conference on Learning Representations },
year={2023},
}

@INPROCEEDINGS{autovitacc,
  author={Li, Zhengang and Sun, Mengshu and Lu, Alec and Ma, Haoyu and Yuan, Geng and Xie, Yanyue and Tang, Hao and Li, Yanyu and Leeser, Miriam and Wang, Zhangyang and Lin, Xue and Fang, Zhenman},
  booktitle={2022 32nd International Conference on Field-Programmable Logic and Applications (FPL)}, 
  title={Auto-ViT-Acc: An FPGA-Aware Automatic Acceleration Framework for Vision Transformer with Mixed-Scheme Quantization}, 
  year={2022},
  volume={},
  number={},
  pages={109-116},
  doi={10.1109/FPL57034.2022.00027}}

@article{drvit,
    title = {{DRViT}: A dynamic redundancy-aware vision transformer accelerator via algorithm and architecture co-design on {FPGA}},
    journal = {Journal of Parallel and Distributed Computing},
    volume = {199},
    pages = {105042},
    year = {2025},    
    author = {Xiangfeng Sun and Yuanting Zhang and Qinyu Wang and Xiaofeng Zou and Yujia Liu and Ziqian Zeng and Huiping Zhuang},   
}

@INPROCEEDINGS{heatvit,
  author={Dong, Peiyan and Sun, Mengshu and Lu, Alec and Xie, Yanyue and Liu, Kenneth and Kong, Zhenglun and Meng, Xin and Li, Zhengang and Lin, Xue and Fang, Zhenman and Wang, Yanzhi},
  booktitle={2023 IEEE International Symposium on High-Performance Computer Architecture (HPCA)}, 
  title={HeatViT: Hardware-Efficient Adaptive Token Pruning for Vision Transformers}, 
  year={2023},
  volume={},
  number={},
  pages={442-455},
  doi={10.1109/HPCA56546.2023.10071047}}

@INPROCEEDINGS{staticanddynamicpruning,
  author={Parikh, Dhruv and Li, Shouyi and Zhang, Bingyi and Kannan, Rajgopal and Busart, Carl and Prasanna, Viktor},
  booktitle={2024 IEEE 32nd Annual International Symposium on Field-Programmable Custom Computing Machines (FCCM)}, 
  title={Accelerating ViT Inference on FPGA through Static and Dynamic Pruning}, 
  year={2024},
  volume={},
  number={},
  pages={78-89},
  doi={10.1109/FCCM60383.2024.00018}}

@INPROCEEDINGS{adaptiv,
  author={Yoo, Seungjae and Kim, Hangyeol and Kim, Joo-Young},
  booktitle={2024 57th IEEE/ACM International Symposium on Microarchitecture (MICRO)}, 
  title={AdapTiV: Sign-Similarity Based Image-Adaptive Token Merging for Vision Transformer Acceleration}, 
  year={2024}, 
  pages={64-77},
  doi={10.1109/MICRO61859.2024.00015}}

@INPROCEEDINGS{vitcod,
  author={You, Haoran and Sun, Zhanyi and Shi, Huihong and Yu, Zhongzhi and Zhao, Yang and Zhang, Yongan and Li, Chaojian and Li, Baopu and Lin, Yingyan},
  booktitle={2023 IEEE International Symposium on High-Performance Computer Architecture (HPCA)}, 
  title={ViTCoD: Vision Transformer Acceleration via Dedicated Algorithm and Accelerator Co-Design}, 
  year={2023},
  volume={},
  number={},
  pages={273-286},
  doi={10.1109/HPCA56546.2023.10071027}}

@ARTICLE{via,
  author={Wang, Teng and Gong, Lei and Wang, Chao and Yang, Yang and Gao, Yingxue and Zhou, Xuehai and Chen, Huaping},
  journal={IEEE Transactions on Computer-Aided Design of Integrated Circuits and Systems}, 
  title={{ViA}: A Novel Vision-Transformer Accelerator Based on {FPGA}}, 
  year={2022},
  volume={41},
  number={11},
  pages={4088-4099},
  doi={10.1109/TCAD.2022.3197489}}

@inproceedings{swat,
    author = {Dong, Qiwei and Xie, Xiaoru and Wang, Zhongfeng},
    title = {{SWAT}: An Efficient Swin Transformer Accelerator Based on {FPGA}},
    year = {2024},       
    doi = {10.1109/ASP-DAC58780.2024.10473931},   
    booktitle = {Proceedings of the 29th Asia and South Pacific Design Automation Conference (ASPDAC)},
    pages = {515--520},
    numpages = {6},    
}

@INPROCEEDINGS{swin,
  author={Liu, Ze and Lin, Yutong and Cao, Yue and Hu, Han and Wei, Yixuan and Zhang, Zheng and Lin, Stephen and Guo, Baining},
  booktitle={2021 IEEE/CVF International Conference on Computer Vision (ICCV)}, 
  title={Swin Transformer: Hierarchical Vision Transformer using Shifted Windows}, 
  year={2021},
  volume={},
  number={},
  pages={9992-10002},
  doi={10.1109/ICCV48922.2021.00986}}

@ARTICLE{outputblock,
  author={Zhao, Zhongyu and Cao, Rujian and Un, Ka-Fai and Yu, Wei-Han and Mak, Pui-In and Martins, Rui P.},
  journal={IEEE Transactions on Circuits and Systems II: Express Briefs}, 
  title={An {FPGA}-Based Transformer Accelerator Using Output Block Stationary Dataflow for Object Recognition Applications}, 
  year={2023},
  volume={70},
  number={1},
  pages={281-285},
  doi={10.1109/TCSII.2022.3196055}}

@INPROCEEDINGS{mevit,
  author={Marino, Kyle and Zhang, Pengmiao and Prasanna, Viktor K.},
  booktitle={2023 IEEE 30th International Conference on High Performance Computing, Data, and Analytics (HiPC)}, 
  title={ME- ViT: A Single-Load Memory-Efficient FPGA Accelerator for Vision Transformers}, 
  year={2023},
  volume={},
  number={},
  pages={213-223},
  doi={10.1109/HiPC58850.2023.00039}}

@ARTICLE{109,
  author={Zhang, Yueqi and Feng, Lichen and Shan, Hongwei and Zhu, Zhangming},
  journal={IEEE Transactions on Circuits and Systems for Video Technology}, 
  title={A 109-GOPs/W FPGA-Based Vision Transformer Accelerator With Weight-Loop Dataflow Featuring Data Reusing and Resource Saving}, 
  year={2024},
  volume={34},
  number={12},
  pages={13596-13610},
  doi={10.1109/TCSVT.2024.3439600}
  }

@INPROCEEDINGS{vita,
  author={Nag, Shashank and Datta, Gourav and Kundu, Souvik and Chandrachoodan, Nitin and Beerel, Peter A.},
  booktitle={2023 IEEE International Symposium on Circuits and Systems (ISCAS)}, 
  title={{ViTA}: A Vision Transformer Inference Accelerator for Edge Applications}, 
  year={2023},  
  pages={1-5},
  doi={10.1109/ISCAS46773.2023.10181988}}

@inproceedings{
lightweight,
title={Lightweight Vision Transformers for Low Energy Edge Inference},
author={Shashank Nag and Logan Liberty and Aishwarya Sivakumar and Neeraja J Yadwadkar and Lizy Kurian John},
booktitle={Machine Learning for Computer Architecture and Systems 2024},
year={2024},
}

@misc{vaqf,
      title={{VAQF}: Fully Automatic Software-Hardware Co-Design Framework for Low-Bit Vision Transformer}, 
      author={Mengshu Sun and Haoyu Ma and Guoliang Kang and Yifan Jiang and Tianlong Chen and Xiaolong Ma and Zhangyang Wang and Yanzhi Wang},
      year={2022},
      eprint={2201.06618},
      archivePrefix={arXiv},
      primaryClass={cs.LG},
      url={https://arxiv.org/abs/2201.06618}, 
}

@InProceedings{efficientvit,
    title     = {EfficientViT: Memory Efficient Vision Transformer with Cascaded Group Attention},
    author    = {Liu, Xinyu and Peng, Houwen and Zheng, Ningxin and Yang, Yuqing and Hu, Han and Yuan, Yixuan},
    booktitle = {Proceedings of the IEEE/CVF Conference on Computer Vision and Pattern Recognition (CVPR)},   
    pages     = {15094--15104},
    year      = {2023}
}

@INPROCEEDINGS{an_energy_efficient,
  author={Cao, Jiacheng and Guo, Jiaqi and Xiong, Wei and Luo, Huanlin and Wang, Jian and Lai, Jinmei},
  booktitle={2025 IEEE 33rd Annual International Symposium on Field-Programmable Custom Computing Machines (FCCM)}, 
  title={An Energy-Efficient FPGA-Based Vision Transformer Accelerator via Software-Hardware Co-Design}, 
  year={2025}, 
  pages={272-272}, 
  doi={10.1109/FCCM62733.2025.00032}
  }

@ARTICLE{m2vit,
  author={Liang, Yanbiao and Shi, Huihong and Wang, Zhongfeng},
  journal={IEEE Transactions on Very Large Scale Integration (VLSI) Systems}, 
  title={M2-ViT: Accelerating Hybrid Vision Transformers With Two-Level Mixed Quantization}, 
  year={2025},
  volume={33},
  number={5},
  pages={1492-1496},
  doi={10.1109/TVLSI.2024.3525184}}

@InProceedings{DeiT,
  title = 	 {Training data-efficient image transformers \& distillation through attention},
  author =       {Touvron, Hugo and Cord, Matthieu and Douze, Matthijs and Massa, Francisco and Sablayrolles, Alexandre and Jegou, Herve},
  booktitle = 	 {Proceedings of the 38th International Conference on Machine Learning},
  pages = 	 {10347--10357},
  year = 	 {2021},
  editor = 	 {Meila, Marina and Zhang, Tong},
  volume = 	 {139},
  series = 	 {Proceedings of Machine Learning Research},
  month = 	 {18--24 Jul},
  publisher =    {PMLR},
}

@inproceedings{ImageNet,
    author={Deng, Jia and Dong, Wei and Socher, Richard and Li, Li-Jia and Kai Li and Li Fei-Fei},
    booktitle={2009 IEEE Conference on Computer Vision and Pattern Recognition},
    title={ImageNet: A large-scale hierarchical image database},
    year={2009},  
    pages={248-255},
    doi={10.1109/CVPR.2009.5206848}
}

@online{makerfile,
    author = {{Makerfocus}},
    title = {{USB Power Meter Digital Display - Voltage Current Amps Capacity Time Temperature Meter}},
    year = {2025}, 
}

@inproceedings{amdahl,
  title={Validity of the single processor approach to achieving large scale computing capabilities},
  author={Amdahl, Gene M},
  booktitle={Proceedings of the April 18-20, 1967, spring joint computer conference},
  pages={483--485},
  year={1967}
}

@inproceedings{10557992,
  author={Shao, Haikuo and Shi, Huihong and Mao, Wendong and Wang, Zhongfeng},
  booktitle={2024 IEEE International Symposium on Circuits and Systems (ISCAS)}, 
  title={An FPGA-Based Reconfigurable Accelerator for Convolution-Transformer Hybrid EfficientViT}, 
  year={2024},
  volume={},
  number={},
  pages={1-5},
  doi={10.1109/ISCAS58744.2024.10557992}}

@article{
vit-tiny,
title={How to train your ViT? Data, Augmentation, and Regularization in Vision Transformers},
author={Andreas Peter Steiner and Alexander Kolesnikov and Xiaohua Zhai and Ross Wightman and Jakob Uszkoreit and Lucas Beyer},
journal={Transactions on Machine Learning Research},
issn={2835-8856},
year={2022},
note={}
}

@article{gibsonDLAS2025,
  title      = {{{DLAS}}: {{A Conceptual Model}} for {{Across-Stack Deep Learning Acceleration}}},
  shorttitle = {{{DLAS}}},
  author     = {Gibson, Perry and Cano, Jose and Crowley, Elliot and Storkey, Amos and O’boyle, Michael},
  year       = {2025},
  journal    = {ACM Transactions on Architecture and Code Optimization (TACO)}
}

\end{document}